\def\year{2021}\relax
\documentclass[letterpaper]{article} 

\usepackage{aaai21}  
\usepackage{times}  
\usepackage{helvet} 
\usepackage{courier}  
\usepackage[hyphens]{url}  
\usepackage{graphicx} 
\usepackage{amsfonts}
\usepackage{amsmath}
\usepackage{booktabs}
\usepackage{multirow}
\usepackage{algorithm}
\usepackage{algorithmic}
\usepackage{subfigure}
\usepackage[switch]{lineno}
\usepackage{graphicx}  
\usepackage{natbib}  
\usepackage{caption} 
\title{Deep Graph-neighbor Coherence Preserving Network for Unsupervised Cross-modal Hashing}
\author {
        Jun Yu\textsuperscript{\rm 1}\footnote{Jun Yu is the corresponding author },
        Hao Zhou\textsuperscript{\rm 1},
        Yibing Zhan\textsuperscript{\rm 1}, 
        Dacheng Tao\textsuperscript{\rm 2}
        \\
}
\affiliations {
    \textsuperscript{\rm 1}  HangZhou Dianzi University \\
    \textsuperscript{\rm 2}  The University of Sydney \\
    yujun@hdu.edu.cn, zhouhao@hdu.edu.cn, zybjy@hdu.edu.cn, dacheng.tao@sydney.edu.au
}

\begin{document} 

\maketitle

\begin{abstract}
Unsupervised cross-modal hashing (UCMH) has become a hot topic recently. Current UCMH focuses on exploring data similarities. However, current UCMH methods calculate the similarity between two data, mainly relying on the two data's cross-modal features. These methods suffer from inaccurate similarity problems that result in a suboptimal retrieval Hamming space, because the cross-modal features between the data are not sufficient to describe the complex data relationships, such as situations where two data have different feature representations but share the inherent concepts. In this paper, we devise a deep graph-neighbor coherence preserving network (DGCPN). Specifically, DGCPN stems from graph models and explores graph-neighbor coherence by consolidating the information between data and their neighbors. DGCPN regulates comprehensive similarity preserving losses by exploiting three types of data similarities (i.e., the graph-neighbor coherence, the coexistent similarity, and the intra- and inter-modality consistency) and designs a half-real and half-binary optimization strategy to reduce the quantization errors during hashing. Essentially, DGCPN addresses the inaccurate similarity problem by exploring and exploiting the data's intrinsic relationships in a graph. We conduct extensive experiments on three public UCMH datasets. The experimental results demonstrate the superiority of DGCPN, e.g., by improving the mean average precision from 0.722 to 0.751 on MIRFlickr-25K using 64-bit hashing codes to retrieve texts from images. We will release the source code package and the trained model on https://github.com/Atmegal/DGCPN.
\end{abstract}
\begin{figure}[t]
\centering
\includegraphics[width=1\linewidth]{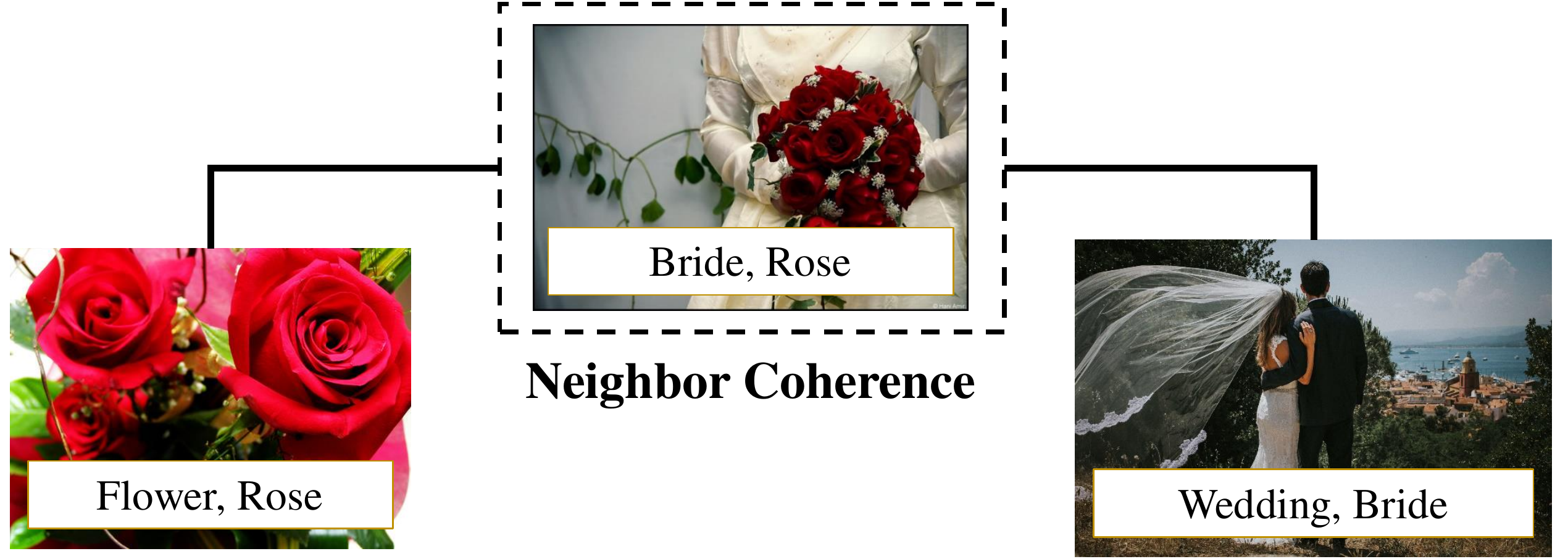}
\caption{Cross-modal examples of three images and their coexistent tags. Rose and wedding have high similarities because the rose is a fundamental component of a wedding. However, this intricate relationship cannot be derived by using visual appearances and linguistic tags of the given ``rose'' and ``wedding''. Nevertheless, such a connection can be inferred by their neighbor coherence that both ``rose'' and ``wedding'' have high similarities of the ``bride, rose".}
\label{fig1}
\end{figure}

\begin{figure}[t]
\centering
\includegraphics[width=1.0\linewidth]{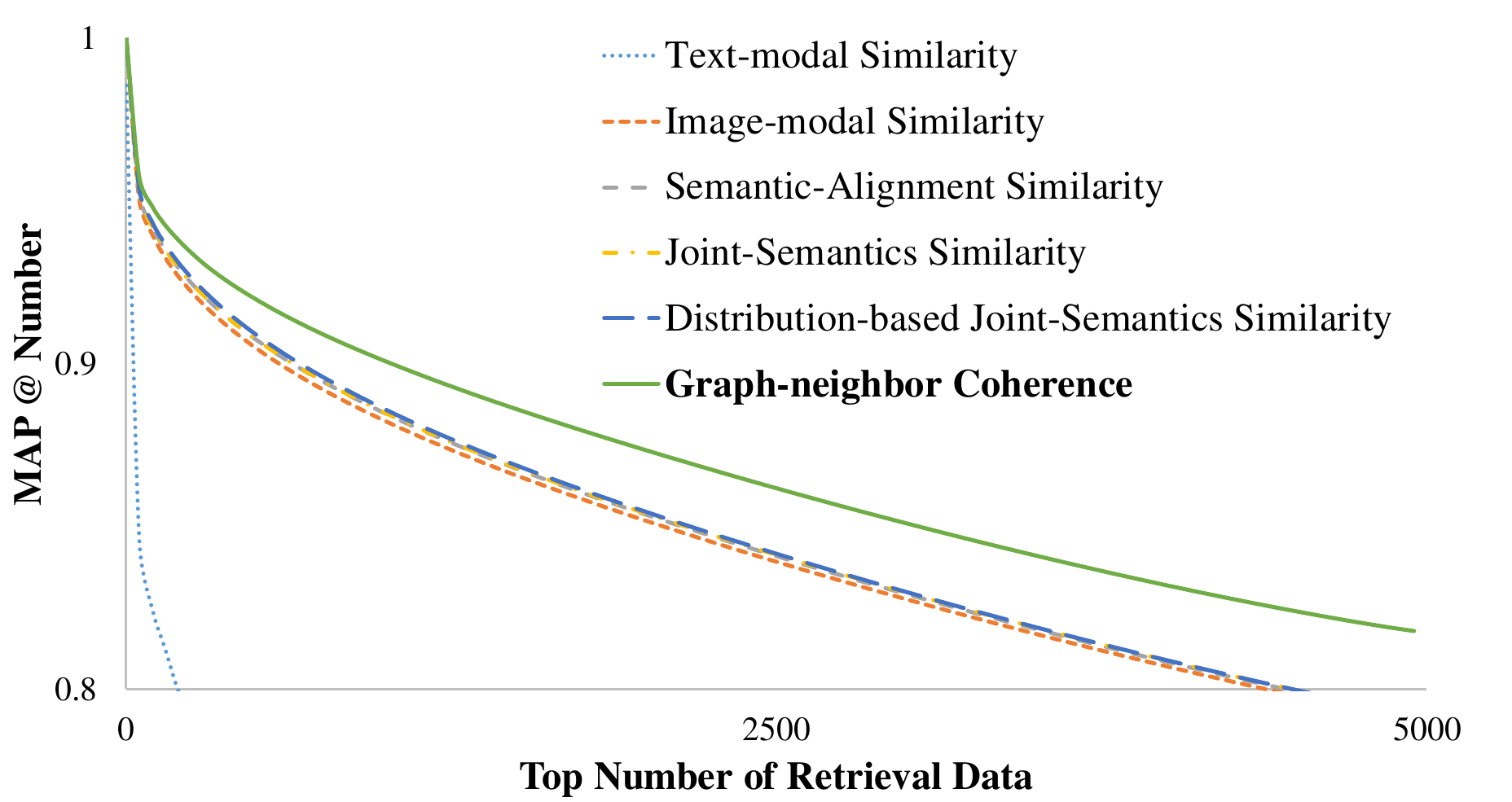}
\caption{Comparison of six types of data similarities using 5000 data on MIRFlickr-25K \cite{huiskes2008mir}. MAP@Number is the mean average precision (MAP) based on the top number of retrieval data. Following \cite{hu2020creating}, we use a pre-trained VGG-19 \cite{simonyan2014very} to extract image features and use the bag-of-words (BoW) model to extract text features. Text-modal and image-modal similarities are cosine similarities using the text and image features. Semantic-alignment similarity \cite{yang2020deep} fuses text and image modal similarities. Both joint-semantics similarity \cite{su2019deep} and distribution-based joint-semantics similarity \cite{yang2020deep} additionally exploit data distributions. Graph-neighbor coherence is the similarity proposed in this paper. We can conclude that graph-neighbor coherence has the best consistency with the real similarities of labels.}
\label{fig2}
\end{figure}

\section{Introduction}
Cross-modal retrieval (CMR) aims to retrieve relevant data when given queries that have different modalities of the retrieval data \cite{wang2016comprehensive}. Due to the explosion of multimedia data on the Internet \cite{yu2013exploiting,yang2020generative}, cross-modal retrieval has become an essential technique in information retrieval \cite{yang2020generative} and multi-modal data mining \cite{qiao2019mirrorgan,qiao2019learn,yang2020joint,yang2020generative,li2020deepfacepencil}. Generally, CMR methods map heterogeneous data into a common space. In this common space, similarities between data can be directly measured. However, most CMR methods use real-valued common spaces, and they suffer from high computation burden for a large volume of multimedia data \cite{feng2014cross,wang2017adversarial,xie2020multi}. By contrast, cross-modal hashing (CMH) methods use compact Hamming space. Due to lower data storage and higher retrieval efficiency, CMH methods have become a hot topic recently.

CMH methods can be roughly classified into two categories: supervised methods \cite{jiang2017deep} and unsupervised methods \cite{su2019deep}. Supervised methods require semantic labels, whereas unsupervised methods only need to know whether heterogeneous data are coexistent \cite{wang2016comprehensive}. In most real-world situations, semantic labels are unavailable, and manual annotations are time-consuming and expensive; and consequently, unsupervised CMH (UCMH) methods are more flexible and feasible.

Current UCMH methods focused on generating data similarities that reflect the data's semantic relationships. They used data similarities as optimization targets to preserve data similarities during hashing \cite{su2019deep,yang2020deep,liu2020joint}. Although current UCMH methods achieved promising improvements, they still suffer from inaccurate similarity problems, limiting retrieval performance. Most methods calculate similarities between data, using features only between the corresponding two data \cite{yang2020deep}. However, features between data are insufficient to describe intricate data relationships; for example, rose and wedding have higher similarities because the rose is the basic component of the wedding. Nevertheless, this relationship cannot be derived from visual appearances and linguistic tags of the given ``rose'' and ``wedding'' in Fig. \ref{fig1}. Intuitively, such high similarity can be inferred from their coherence of neighbors, i.e., both ``rose'' and ``wedding'' have high similarities of the ``bride and rose." 

Even though some methods attempt to consider data distribution \cite{su2019deep,liu2020joint},  the similarities of these methods are calculated, mainly relying on the features between the data; their similarities still have huge gaps compared with the real situations derived from labels. Fig. \ref{fig2} compares the mean average precision (MAP) of six types of data similarities using 5000 data of MIRFlickr-25K \cite{huiskes2008mir}. Text-modal and image-modal similarities are obtained using single modal features. Semantic-alignment similarity \cite{yang2020deep} fuses text and image modal similarities. Both joint-semantics similarity \cite{su2019deep} and distribution-based joint-semantics similarity \cite{liu2020joint} consider data distributions. Graph-neighbor coherence is the similarity proposed in this paper. We observe that previous data similarities only slightly outperform the image-model similarities.

In light of the above analysis, we develop a deep graph-neighbor coherence preserving network (DGCPN) for UCMH that has the following main contributions:

\begin{itemize}
\item We propose graph-neighbor coherence (GC) that explores the relationships between data and their neighbors. Specifically, we model the retrieval data as a graph in which each node consists of a group of coexistent data; naturally, the similarities between data are equal to the similarities between the corresponding nodes. The GC is calculated based on a designed conditional probability model that sums all probabilities of whether both two calculating nodes have high similarities of one of their common neighbors. In this manner, GC considers not only the number of common neighbors but also the similarity degree of the neighbors between the two nodes. Here, we suppose that each node is connected to $k$-nearest neighbors ($k$-NN) and obtain the two nodes' probabilities with high similarities using the two nodes' cross-modal features. GC further adds the distance between nodes for robust performance. As shown in Fig. \ref{fig2}, GC effectively reflects the data retrieval relationships and thus addresses the inaccurate similarity problem.

\item DGCPN designs comprehensive similarity preserving losses to regulate three types of similarities: 1) graph-neighbor coherence between data in a subset, 2) coexistent similarity between coexistent data, and 3) intra- and inter-modality consistency between similarities obtained from different modalities. The comprehensive similarity preservation guarantees the robust learning process, and the intrinsic relationships of data (i.e., the coexistent similarity and the intra- and inter-modality consistency) alleviate the side effects of graph-neighbor coherence's possible inaccurate similarities. 

\item Moreover, DGCPN develops a half-real and half-binary optimization strategy to reduce the value gaps and the similarity gaps between the real-valued space and Hamming space by calculating the losses between one real value and one binary code. The optimization strategy effectively reduces quantization errors during hashing.
\end{itemize}

We conducted extensive experiments on three public datasets. The improved performance compared with the state-of-the-art demonstrates the competitiveness of DGCPN. The rest of the paper is organized as follows: we first review related works. Then, we explain the details of DGCPN and present the experimental results. The last section concludes this paper.

\section{Related UCMH Work}
Previous UCMH methods focused on obtaining binary codes of data and learned the hashing functions by mapping data features into the corresponding binary codes \cite{liong2018cross}. For example, IMH \cite{song2013inter}, CMFH \cite{ding2014collective}, and  MSFH \cite{fang2019unsupervised} learned hashing codes by preserving intra- and inter-graph consistency, using collective matrix factorization, and using multigraph regularized smooth matrix factorization, respectively. However, the above methods were space-consuming because obtaining robust binary codes requires considering all training data simultaneously. Besides, most previous methods used human-crafted features and shallow models that limited the retrieval performance.

Later, UCMH methods exploited deep neural networks (DNNs)  \cite{wang2020unsupervised,zhang2020empowering}. For example, \citeauthor{hu2018deep} designed a deep autoencoder with an adjusted Tanh function. \citeauthor{li2019coupled} and \citeauthor{zhang2019multi} proposed coupled cycle generative adversarial networks (GANs) and multi-pathway GANs, respectively. Some UCMH methods learned real values with approximate binary representations and obtained the binary codes from the real values using a Sign function \cite{kumar2011learning}. Most deep methods adopted the above strategy to implement batch-wise training to save time and space \cite{wang2020unsupervised,hu2020creating}.

In UCMH, data similarities serve as critical considerations \cite{liang2016self,li2017multiview}. Earlier UCMH methods calculated data similarities by using discrete models \cite{song2013inter,ding2014collective}. For instance, data pairs with smaller and larger feature distances were set to 1 and -1, respectively \cite{kumar2011learning}. However, discrete models ignore the data pairs with intermediate distances. Later, continuous similarities between data were designed by using feature distances \cite{xie2016unsupervised,fang2019unsupervised}. Nevertheless, the data similarities have been generally used as weights in the optimization functions and did not make full use of data similarities \cite{wu2018unsupervised,wang2020unsupervised,hu2020creating}. Most current methods learned retrieval Hamming space by preserving calculated data similarities. \citeauthor{su2019deep}, \citeauthor{yang2020deep}, and \citeauthor{liu2020joint} adopted semantic-alignment similarities fusing cross-modal information. Moreover, \citeauthor{su2019deep} and \citeauthor{liu2020joint} considered structural information. However, recent UCHM methods still suffer from inaccurate similarity problems and obtain suboptimal retrieval Hamming space.

\begin{figure*}[t]

\centerline{\includegraphics[width=0.83\linewidth]{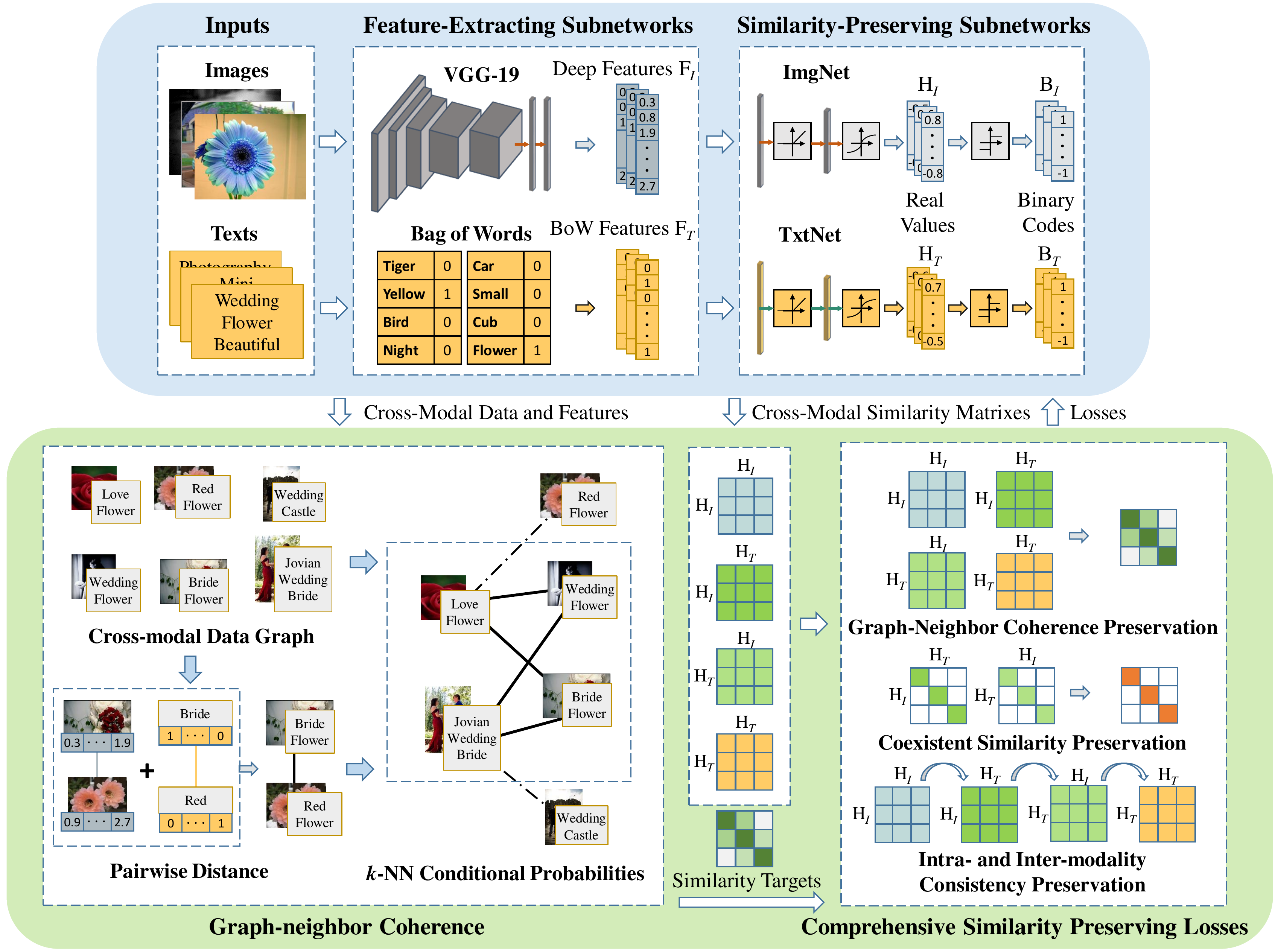}}
\caption{The framework of DGCPN. The blue background illuminates the network details. Specifically, DGCPN first extracts image and text features by using feature-extracting subnetworks. Afterward, DGCPN maps the image and text features into relaxed real values and binary codes using similarity-preserving subnetworks. The green background shows additional training details. During training, the graph-neighbor coherence of training data are calculated by analyzing cross-modal data features. Then, DGCPN is trained by calculating the comprehensive similarity preserving losses based on a batch size of data.}
\label{fig3}
\end{figure*}

\section{DGCPN}
This section presents the proposed DGCPN. Specifically, we first introduce the framework and preliminary definitions. We then explain the details of the graph-neighbor coherence, the comprehensive similarity preserving losses, and the half-real and half-binary optimization strategy sequentially.

\subsection{The Framework and Preliminary Definitions}
Fig. \ref{fig3} presents the framework of DGCPN. Without loss of generality, we use texts and images as the cross-modal examples. The blue background illuminates the network details. Specifically, DGCPN consists of two sequential subnetworks: feature-extracting subnetworks to extract the image and text features and similarity-preserving subnetworks to map the image and text features into binary codes. We note that the feature-extracting subnetworks can marry other backbones, such as AlexNet \cite{krizhevsky2012imagenet} and the Latent Dirichlet allocation (LDA) model \cite{blei2003latent}. The green background shows the training details. Specifically, the GC between data is first obtained by analyzing the whole training data. Then, DGCPN is trained in a batch-wise manner based on the designed comprehensive similarity preserving losses. Moreover, we exploit a half-real and half-binary optimization strategy to further improve retrieval performance. 

Suppose $(v_i,t_i)$ indicate the $i$-th coexistent image and text. $i\in [M]$. We define $[M]$=$\{1,2...M\}$. $M$ is the total number of coexistent pairs. The features, relaxed real values, and binary codes of the $i$-th image and text are defined as $\vec{f}_{*_i}$ $\in$$ \mathbb{R}^{d_* \times 1}$, $\vec{h}_{*_i}$ $\in$$ \mathbb{R}^{ d_b\times 1}$, and $\vec{b}_{*_i}$$\in \mathbb{R}^{d_b\times 1}$, respectively. $*\in \{I, T\}$. $d_I$, $d_T$, and $d_b$ are the dimension of image features, text features, and hashing coding, respectively. $\vec{b}_{*_i}$ = $\mathrm{sign}(\vec{h}_{*_i})$. The $\mathrm{sign}(.)$ is the Sign function. DGCPN adopts batch-wise training. Suppose $\textbf{F}_*$=$\{\vec{f}_{*_i}$$,i\in[N]\}$$\in$$\mathbb{R}^{ d_*\times N}$, $\textbf{H}_*$=$\{\vec{h}_{*_i}$$,i\in[N]\}$$\in$$\mathbb{R}^{ d_b\times N}$, and $\textbf{B}_*$=$\{\vec{b}_{*_i},i\in[N]\}$$\in \mathbb{R}^{ d_b\times N}$ indicate a set of features, relaxed real values, and binary codes with the batchsize of $N$, respectively. Besides, 
\begin{equation}
\textbf{H}_I=\mathrm{ImgNet}(\textbf{F}_I, \theta_I)\ \mathrm{and}\  \textbf{H}_T=\mathrm{TxtNet}(\textbf{F}_T, \theta_T),
\end{equation}
where $\theta_I$ and $\theta_T$ are training parameters.  

We use cosine similarity to measure the similarity (or distance) between vectors. The cosine similarity is defined as:
\begin{equation}
\mathrm{c}(\vec{x}_i, \vec{x}_j)=\frac{\vec{x}_i^T\vec{x}_j}{\|\vec{x}_i\|_2\|\vec{x}_j\|_2},
\end{equation}
where $\vec{x}_i^T$ is the transpose of $\vec{x}_i$. $\|.\|_2$ is the L2-norm of vectors. We select cosine similarity because cosine similarity is commonly adopted in deep UCMH methods \cite{yang2020deep,liu2020joint}. The cosine similarities between two sets of vectors are defined as follows:
\begin{equation}
\mathcal{C}(\textbf{X}, \textbf{X})=
\left[
 \begin{array}{cccc} 
   \mathrm{c}(\vec{x}_1,\vec{x}_1) & \mathrm{c}(\vec{x}_1,\vec{x}_2) &\cdots& \mathrm{c}(\vec{x}_1,\vec{x}_N)\\
   \mathrm{c}(\vec{x}_2,\vec{x}_1) & \mathrm{c}(\vec{x}_2,\vec{x}_2) &\cdots& \mathrm{c}(\vec{x}_2,\vec{x}_N)\\
\vdots&\vdots&\ddots&\vdots\\
\mathrm{c}(\vec{x}_N,\vec{x}_1) & \mathrm{c}(\vec{x}_N,\vec{x}_2) &\cdots& \mathrm{c}(\vec{x}_N,\vec{x}_N)\\
  \end{array}
  \right],
\end{equation}
where $\textbf{X}$=$\{\vec{x}_{_i},i\in[N]\}$. 

\subsection{Graph-neighbor Coherence}
As aforementioned, previous methods generally calculated similarities between data, mainly relying on the two data's cross-modal features; and current data similarities still have huge gaps compared with the real similarities obtained from labels. DGCPN proposes graph-neighbor coherence (GC) that improves the accuracy of previous data similarities by combining cross-modal features between two data and relationships between the two data and their neighbors. This subsection introduces the details of GC.

Specifically, we model all of the training data as a graph $\mathcal{G}=\{\mathcal{O},\mathcal{E}\}$. $\mathcal{O}=\{o_i, i\in [M]\}$. $o_i=(v_i, t_i)$. Suppose $l_i$ indicates the label of $v_i$, $t_i$, and $o_i$. Therefore, the problem of calculating similarities between the data is transformed into calculating similarities between the nodes. We model GC between the nodes as a conditional probability problem:
\begin{equation} \label{eq4}
P(l_i=l_j)=P(l_i=l_j|\mathcal{O}).
\end{equation}

However, the calculation of GC based on Eq. (\ref{eq4}) is complicated because there are $ M $ joint variables that constitute numerous conditional combinations. For computational efficiency, we impose the limitation in calculating GC between two nodes that only one other node is used as a condition. According to the total probability rule, Eq. (\ref{eq4}) is revised as:
\begin{equation} \label{eq5}
\begin{split}
P&(l_i=l_j)\\
&=\sum_{q=1}^MP(l^F_i=l^F_{q}|o_i,o_q)P(l^F_j=l^F_q|o_j,o_q),\\
\end{split}
\end{equation}
where $l^F_*$ is the virtual label of $o_*$. We use virtual labels to discuss the similarities between nodes. $P(l^F_i=l^F_q|o_i,o_q)$ indicates the probability that $o_i$ and $o_q$ have the same virtual labels or have higher similarities. We suppose each node is related to its $k$-nearest neighbors and following \citeauthor{zhai2013learning,zhan2018comprehensive} define
 $P(l^F_i=l^F_q|o_i,o_q)$ as:
\begin{equation}\label{eq6}
\begin{split}
P&(l^F_i=l^F_q|o_i,o_q) \\
&=\left\{
\begin{array}{ll} 
\frac{\mathrm{d}(o_i,o_q)}{\sum_{p\in \mathbf{Ne}(o_i,k)}\mathrm{d}(o_i,o_p)} & {o_q \in \mathbf{Ne}(o_i,k)}\\ 
0 & \mathrm{else}
\end{array} \right.
\end{split},
\end{equation}
where $\mathrm{d}(o_i,o_q)$ is the pairwise distance between $o_i$ and $o_q$:
\begin{equation} \label{eq7}
\mathrm{d}(o_i,o_q)=(1-\alpha )\mathrm{c}(\vec{f}_{I_i},\vec{f}_{I_q}) + \alpha \mathrm{c}(\vec{f}_{T_i},\vec{f}_{T_q}),
\end{equation}
and $\mathbf{Ne}(o_i,k)$ indicates the set of $k$-nearest neighbors of $o_i$ using Eq. (\ref{eq7}). Note that $\mathrm{c}(\vec{f}_{I_i},\vec{f}_{I_q})$, $\mathrm{c}(\vec{f}_{T_i},\vec{f}_{T_q})$, and $\mathrm{d}(o_i,o_q)$ are in the 0-1 range because of the adopted deep features and BoW features. The graph-neighbor coherence between $o_i$ and $o_j$ is calculated as:
\begin{equation}\label{eq8}
s(o_i,o_j) =  (1-\gamma)\mathrm{d}(o_i,o_j)  + \gamma \beta P(l_i=l_j), 
\end{equation}
where $\beta$ is used to adjust the range of $P(l_i=l_j)$. $\gamma$ is a trade off parameter. We add $\mathrm{d}(o_i,o_j)$ for robust performance. We note that the cosine similarity values range between -1 to 1. Therefore, following \citeauthor{liu2020joint}, we use 2$s(o_i,o_k)$-1 as the final graph-neighbor coherence during training.

\subsection{Comprehensive Similarity Preserving Losses}
Current methods have mainly focused on preserving similarities and rarely proposed solutions to solve the inaccurate similarity problem. By contrast, DGCPN proposes comprehensive similarity preserving losses by maintaining data similarities from three complementary aspects: graph-coherence preservation, coexistent similarity preservation, and intra- and inter-modality consistency preservation. The three types of complementary similarity preservation guarantee the robust similarity learning process. The exploitation of the intrinsic relationships between the data (i.e., the coexistent similarity and the intra- and inter-modality consistency) alleviates the side effect of inaccurate similarities.
 
Suppose that there exists a batch size of training data with relaxed real values: $ \textbf{H}_I$ and $ \textbf{H}_T$. Then, four types of similarity matrixes are generated, namely, two homogeneous data similarity matrixes: $\mathcal{C}(\textbf{H}_I, \textbf{H}_I)$, $\mathcal{C}(\textbf{H}_T, \textbf{H}_T)$ and two heterogeneous data similarities matrixes: $\mathcal{C}(\textbf{H}_I, \textbf{H}_T)$, $\mathcal{C}(\textbf{H}_T, \textbf{H}_I)$. 

The graph-neighbor coherence preserving losses $L_{g}(\textbf{H}_I, \textbf{H}_T) $, the coexistent similarity preserving losses $L_{c}(\textbf{H}_I, \textbf{H}_T)$, and the intra- and inter-modality consistency preserving losses $L_{i}(\textbf{H}_I, \textbf{H}_T)$ are respectively defined as:
\begin{equation}
L_{g}(\textbf{H}_I, \textbf{H}_T) = \sum_{p,q}
\|\mathcal{C}(\textbf{H}_p, \textbf{H}_q) - S_{gc}(\textbf{H}_I, \textbf{H}_T)\|_F,
\end{equation} 
\begin{equation}
\label{eq12}
L_{c}(\textbf{H}_I, \textbf{H}_T) =  \|\mathrm{Tr}(\mathcal{C}(\textbf{H}_I, \textbf{H}_T) - 1.5\mathbf{I} )\|_2,
\end{equation} 
\begin{equation}
L_{i}(\textbf{H}_I, \textbf{H}_T) = \sum_{p,q,p_1, q_1}
\|\mathcal{C}(\textbf{H}_p, \textbf{H}_q) - \mathcal{C}(\textbf{H}_{p_1}, \textbf{H}_{q_1}) \|_F,
\end{equation}
where $p,q,p_1,q_1\in\{I,T\}$. $S_{gc}(\textbf{H}_I, \textbf{H}_T)$ is the corresponding GC of $\textbf{H}_I, \textbf{H}_T$. $\|.\|_F$ is the Frobenius norm. 
$\mathbf{I}$ is an identity matrix  with the same size of $\mathcal{C}(\textbf{H}_I, \textbf{H}_T)$. Following \citeauthor{liu2020joint}, we use 1.5 as the optimization goal of the coexistent similarity preserving that slightly improves performance.
$\mathrm{Tr(.)}$ is the matrix trace. For simplicity, we set all similarity matrixes to have the same significance.

Graph-neighbor coherence reflects the similarities of whether two data share the same labels and is used as the supervised information in $L_{g}(\textbf{H}_I, \textbf{H}_T) $. However, the GC may still obtain errors. Intuitively, the coexistent similarities between coexistent data should be high; and the intra- and inter-modality consistency represents that data's similarities should equal the corresponding nodes' similarities, regardless of the data's modalities. Both the coexistent similarities and intra- and inter-modality consistency are intrinsic data relationships. Therefore, the exploitation of the coexistent similarity and intra- and inter-modality consistency alleviates the side effect of GC's possible errors.

The final similarity preserving losses are calculated as:
\begin{equation}
\begin{split}
L&(\textbf{H}_I, \textbf{H}_T)=\\
&L_{c}(\textbf{H}_I, \textbf{H}_T) + \lambda _1 L_{g}(\textbf{H}_I, \textbf{H}_T) + \lambda _2 L_{i}(\textbf{H}_I, \textbf{H}_T),
\end{split}
\end{equation}
where $\lambda _1$ and $\lambda _2$ are the parameters used to adjust the relative significance of the three similarity preserving losses.

\begin{algorithm}[t] 
\caption{Graph-neighbor Coherence Preserving} 
\label{alg:Framwork} 
\begin{algorithmic}[1] 
\REQUIRE ~~\\ 
$M$ Training Images and Texts; Validation Images and Texts; Batch size $N$; hash code length $d_b$; Max training epoch $E$; trade-off parameters $\alpha$, $\gamma$, $\lambda _1$, and $\lambda _2$; $k$-nearest number and scale parameter $\beta$;\\
\ENSURE ~~\\ 
Hashing Function $\mathrm{ImgNet}(., \theta_I)\ \mathrm{and}\  \mathrm{TxtNet}(., \theta_T)$;\\
\STATE Initial $\theta_I$ and $\theta_T$; 
\STATE Extract image and text features of Training set and obtain graph-neighbor coherence of all training data;\\
\FOR{each $i \in [1,E]$}
\FOR{each $j \in [1,M/N]$}
\STATE obtain training data with batch size of $N$ and the corresponding $S_{gc}(\textbf{H}_I, \textbf{H}_T)$;\\
\STATE update $\theta_I$ and $\theta_T$ using $L(\textbf{H}_I, \textbf{H}_T)$; 
\STATE update $\theta_I$ using $L(\textbf{H}_I, \textbf{B}_T)$;
\STATE update $\theta_T$ using $L(\textbf{B}_I, \textbf{H}_T)$;
\ENDFOR
\STATE calculate MAP of Validation set; if convergence, stop;
\ENDFOR
\RETURN $\mathrm{ImgNet}(., \theta_I)\ \mathrm{and}\  \mathrm{TxtNet}(., \theta_T)$;
\end{algorithmic}
\end{algorithm}

\subsection{Half-real and Half-binary Optimization Strategy}
Previous methods only reduce the value gaps between relaxed real values and binary codes. They still ignore the similarity gaps between the real-valued space and Hamming space. DGCPN proposes a half-real and half-binary optimization strategy to reduce both value gaps and similarity gaps. Intuitively, DGCPN reduces both gaps by additionally preserving similarities in the Hamming space. However, learning with pure binary codes is difficult. Therefore, we relax this condition by calculating similarities between one real value and one binary code. Specifically, we use $L(\textbf{H}_I, \textbf{H}_T)$ to train the whole networks. Then, we use $L(\textbf{H}_I, \textbf{B}_T)$ and $L(\textbf{B}_I, \textbf{H}_T)$ to train the networks of images and texts, respectively. The details of the optimization are presented in the Algorithm \ref{alg:Framwork}.

\begin{table*}[t]
\centering
\begin{small}
 \begin{tabular}{llcccccccccc}
    \toprule
&  & \multicolumn{3}{c}{Wikipedia}&
\multicolumn{3}{c}{MIRFlickr-25K}&
\multicolumn{3}{c}{NUS-WIDE}\\
	\midrule
Task  &Method &	16-bit&	32-bit&	64-bit&	16-bit&	32-bit&	64-bit&	16-bit&	32-bit&	64-bit\\
	\midrule
\multirow{10}{*}{I2T}	
&CVH \cite{kumar2011learning}&0.146&		0.149& 	0.155& 	0.580& 	0.579& 	0.579& 	0.379& 	0.378& 	0.377 \\
&FSH \cite{liu2017cross}&	0.242& 	0.283& 	0.305& 	0.590& 	0.597& 	0.597& 	0.400& 	0.414& 	0.424 \\
&CMFH \cite{ding2014collective}&	0.163& 	0.176& 	0.163& 	0.588& 	0.592& 	0.594& 	0.483& 	0.488& 	0.486 \\ 
&LSSH \cite{zhou2014latent}&	0.364& 	0.387& 	0.397& 	0.630& 	0.634& 	0.631& 	0.475& 	0.484& 	0.474 \\ 
&UGACH \cite{Zhang2018ga}&	0.319& 	0.358& 	0.377& 	0.686& 	0.695& 	0.702& 	0.568& 	0.583& 	0.589 \\
&DJSRH \cite{su2019deep}&	0.384 & 	0.398& 	0.406& 	0.666& 	0.678& 	0.699& 	0.513& 	0.535& 	0.566 \\
&UKD-SS \cite{hu2020creating}&	0.332& 	0.344& 	0.337& 	0.700& 	0.706& 	0.709& 	0.584& 	0.578& 	0.586 \\
&DSAH \cite{yang2020deep}&	0.395& 	0.409& 	0.413& 	0.701& 	0.712& 	0.722& 	0.602& 	0.612& 	0.632 \\
&JDSH \cite{liu2020joint}&	0.351& 	0.383& 	0.399& 	0.669& 	0.683& 	0.698& 	0.554& 	0.561& 	0.582 \\
	\midrule
	&\textbf{DGCPN}&	\textbf{0.404} &	\textbf{0.413} &	\textbf{0.420}& 	\textbf{0.732}& 	\textbf{0.742}& 	\textbf{0.751}& 	\textbf{0.625}& 	\textbf{0.635}& 	\textbf{0.654} \\
	\midrule
	\midrule	
\multirow{10}{*}{T2I}
&CVH \cite{kumar2011learning}&	0.275& 	0.236& 	0.186& 	0.580& 	0.579& 	0.580& 	0.378& 	0.378& 	0.379 \\
&FSH \cite{liu2017cross}&	0.367& 	0.448& 	0.490& 	0.589& 	0.595& 	0.595& 	0.395& 	0.408& 	0.417 \\
&CMFH \cite{ding2014collective}&	0.495& 	0.513& 	0.533& 	0.590& 	0.595& 	0.598& 	0.487& 	0.488& 	0.493 \\ 
&LSSH \cite{zhou2014latent}&	0.403& 	0.411& 	0.422& 	0.621& 	0.628& 	0.626& 	0.476& 	0.481& 	0.477 \\ 
&UGACH \cite{Zhang2018ga}&	0.339& 	0.398& 	0.418& 	0.692& 	0.698& 	0.699& 	0.557& 	0.562& 	0.580 \\ 
&DJSRH \cite{su2019deep}&0.512& 	0.536& 	0.544& 	0.683& 	0.694& 	0.717& 	0.546& 	0.568& 	0.599 \\
&UKD-SS \cite{hu2020creating}&	0.400& 	0.402& 	0.424& 	0.704& 	0.705& 	0.714& 	0.587& 	0.599& 	0.599 \\
&DSAH \cite{yang2020deep}&	0.527& 	0.540& 	0.544& 	0.707& 	0.713& 	0.728& 	0.621& 	0.632& 	0.646 \\
&JDSH \cite{liu2020joint}&	0.398& 	0.448& 	0.480& 	0.686& 	0.699& 	0.716& 	0.580& 	0.596& 	0.626 \\
		\midrule
&\textbf{DGCPN}&	\textbf{0.539}  &	\textbf{0.550} & 	\textbf{0.558} & 	\textbf{0.729} & 	\textbf{0.741} & 	\textbf{0.749} & 	\textbf{0.631} & 	\textbf{0.648} & 	\textbf{0.660} \\
\bottomrule
\end{tabular}
\end{small}
\caption{Performance comparison of ten UCMH methods on three public datasets.}
\label{table1}
\end{table*}
\section{Experiments}

\subsection{Datasets and Evaluation Metrics}
Three public datasets are adopted in our experiments: Wikipedia \cite{rasiwasia2010new}, MIRFlickr-25K \cite{huiskes2008mir}, and NUS-WIDE \cite{chua2009nus}. 

Wikipedia dataset consists of 2,866 image-text pairs from 10 categories. We split Wikipedia dataset into the retrieval/test query/validation query set with 2173/462/231 image-text pairs. The whole retrieval set is used for training. 

MIRFlickr-25K dataset contains 20,015 image-tag pairs with multi labels from 24 classes.  We split MIRFlickr-25K dataset into the retrieval/test query/validation query set with 16,015/2000/2000 image-tag pairs. 5000 image-tag pairs of the retrieval set are used for training.

NUS-WIDE dataset provides 186,577 image-tag pairs with the top-10 concepts. We split NUS-WIDE into retrieval/test query/validation query set with 182,577/2000/2000 image-tag pairs.  5000 image-tag pairs of the retrieval set are used for training. Besides, we select 10,000 image-text pairs from the whole retrieval set as the validation retrieval set for computational efficiency. 

Following \citeauthor{hu2020creating}, for all datasets, the images are represented by 4096-dimensional deep features extracted from VGG-19 pretrained on ImageNet \cite{deng2009imagenet}. The text representations of Wikipedia, MIRFlickr-25K, and NUS-WIDE are 10-dimensional LDA, 1386-dimensional BoW, and 1000-dimensional BoW features, respectively. 

Two cross-modal retrieval tasks are used for testing: text retrieval from an image query (I2T) and image retrieval from a text query (T2I). The retrieval performance is evaluated using MAP. Given one query and the first $R$ top ranked retrieved data, the average precision (AP) is defined as:
\begin{equation}
\mathrm{AP}= \frac{\sum_{q=1}^R P(q)rel(q)}{\sum_{q=1}^R rel(q)},
\end{equation}
where $rel(q)$=1 if the item at rank $q$ is relevant, $rel(q)$=0 otherwise. $P(q)$ denotes the precision of the result ranked at $q$. The MAP is obtained by averaging the AP of all the queries. We report MAP using all retrieval data. 

\subsection{Implementation Details}
The layers of similarity-preserving subnetworks are set as $d_I$-4096-$d_b$ for images and $d_T$-4096-$d_b$ for texts. We fix the feature-extracting subnetworks' parameters during training and only update the parameters of similarity-preserving subnetworks. We use a mini-batch SGD optimizer with a 0.9 momentum and 0.0005 weight decay. The mini-batch size is set to 32. The learning rate is set to 0.005. For simplicity, we perform a 3-step grid search for the parameters. First, we decide the parameter of the pairwise distance. We set $\gamma$=0,  $\lambda _1$=$\lambda _2$=1, and tune $\alpha$ from 0.01, 0.99, and 0.1 to 0.9 at an increment of 0.1 per step. Then, we search for the parameters of GC. $\lambda _1$=$\lambda _2$=1. The value of $\alpha$ is fixed according to the value in the first step. For Wikipedia, we tune both $k$ and $\beta$ from 300 to 1200 at an increment of 300 and $\gamma$ from 0.01, 0.99, 0.1 to 0.9 at an increment of 0.1 per step. For MIRFlickr-25K and NUS-WIDE, we tune $k$ from 500 to 2000 at an increment of 500,  $\beta$ from 2000 to 4500 at an increment of 500, and $\gamma$ from 0.01, 0.99, and 0.1 to 0.9 at an increment of 0.1 per step, respectively. We last tune both $\lambda _1$ and $\lambda _2$ using the grid search from 0.01 to 10 at a product of 10 per step. We use validation sets with early stopping to determine the parameters. All parameters are determined using 64-bit binary codes. The final parameters are: for Wikipedia: $\alpha$=0.3, $\gamma$=0.3, $\lambda _1$=1, $\lambda _2$=1, $\beta$=900, and $k$=600; for MIRFlickr-25K $\alpha$=0.01, $\gamma$=0.3, $\lambda _1$=1, $\lambda _2$=1, $\beta$=4000, $k$=2000; and for NUS-WIDE: $\alpha$=0.1, $\gamma$=0.3, $\lambda _1$=1, $\lambda _2$=1, $\beta$=4500, $k$=2000.

\subsection{Performance Comparison}
We compare DGCPN with nine UCMH methods: CVH \cite{kumar2011learning}, FSH \cite{liu2017cross}, CMFH \cite{ding2014collective}, LSSH \cite{zhou2014latent}, UGACH \cite{Zhang2018ga}, DJSRH \cite{su2019deep}, UKD-SS \cite{hu2020creating}, DSAH \cite{yang2020deep}, and JDSH \cite{liu2020joint}. UGACH, DJSRH, UKD-SS, DSAH, and JDSH are deep models. Besides, DJSRH, DASH, and JDSH explored data similarities. 

We report the MAP of all comparing methods in Table \ref{table1}. DJSRH, DASH, and JDSH were trained in end-to-end training, whereas the rest of the methods were trained using two-step training, which first extracts cross-modal features and then uses fixed features to train the remaining networks. The best performance in Table \ref{table1} is highlighted in bold. An examination of Table \ref{table1} shows that DGCPN obtains the best performance on all datasets regardless of the criteria. In particular, DGCPN outperforms the second-best competitor, DSAH, in MIRFlickr-25K by 4.4\%, 4.2\%, and 4.0\% on 16-bit, 32-bit, and 64-bit for I2T and by 3.1\%, 3.9\%, and 2.9\% on 16-bit, 32-bit, and 64-bit for T2I, respectively.  In NUS-WIDE, compared with DSAH, DGCPN obtains improvements of 3.8\%, 3.8\%, and 3.5\% on 16-bit, 32-bit, and 64-bit for I2T and of 1.6\%, 2.5\%, and 2.2\% on 16-bit, 32-bit, and 64-bit for T2I, respectively. These improvements demonstrate the ability of DGCPN for UCMH. 

\begin{table}[t]
\centering
\begin{small}
\begin{tabular}{lcccc}
\toprule
& \multicolumn{2}{c}{MIRFlickr-25K}& \multicolumn{2}{c}{NUS-WIDE}\\
\midrule
&	I2T &T2I    &	I2T &   T2I\\
	\midrule
\multicolumn{5}{l}{Data Similarity Types}\\
DGCPN-OlPD&   0.725& 0.724 & 	0.626&  0.632 \\
DGCPN-NoPD&	0.748& 	0.748& 	0.651& 	0.656 \\
\midrule
\multicolumn{5}{l}{Similarity Preserving Losses}\\	
DGCPN-GL&	0.734& 0.727&	 0.631& 	0.636\\
DGCPN-GL+CL&	0.746& 	0.749& 	0.649& 0.660 \\
\midrule
\multicolumn{5}{l}{Hashing optimization strategies}\\
DGCPN-Nhash&	0.746& 	0.731& 	0.651& 	0.655 \\
DGCPN-HashL&	0.748& 	0.742& 	0.653& 	0.658 \\
DGCPN-Atanh&	0.746& 	0.739& 	0.656& 	0.662 \\
	\midrule
DGCPN &0.751&   0.749  &0.658&0.663\\
\bottomrule
\end{tabular}
\end{small}
\caption{Ablations experiments on MIRFlickr-25K and NUS-WIDE.}
\label{table2}
\end{table}

\subsection{Ablation Experiments}
In table \ref{table2}, we conduct three types of ablation experiments to discuss the methodological details of DGCPN. For simplicity, we use MIRFlickr-25K and NUS-WIDE as testbeds. 

First, we demonstrate the significance of GC for DGCPN. Two variants are proposed. DGCPN-OlPD is the DGCPN only exploits pairwise distances, and DGCPN-NoPD is the DGCPN does not add pairwise distances. An examination of the results presented in Table \ref{table2} shows that DGCPN-NoPD outperforms DGCPN-OlPD. The improvements indicate that GC is the key success of DGCPN. DGCPN performs the best, which reveals that adding pairwise distances contributes to robust data similarities.

Then, we verify the usefulness of comprehensive similarity preserving losses. Two variants are developed. DGCPN-GL is the DGCPN that only uses graph-coherence preserving loesses, and DGCPN-GL+CL is DGCPN without preserving intra- and inter-modality consistency preserving losses. It is observed that DGCPN-GL shows the worst performance, and DGCPN obtains the best performance. These results demonstrate that exploiting coexistent similarities and intra- and inter-modality consistency alleviates GC's errors and benefits to the retrieval performance.

Last, we show the necessity of the proposed half-real and half-binary optimization strategy. Three variants are designed for comparison. DGCPN-Nhash is DGCPN without using the half-real and half-binary optimization strategy. DGCPN-Atanh and DGCPN-HashL are DGCPN-Nhash that exploits adjusted Tanh \cite{yang2020deep} and a hashing function: $\|\textbf{H}-\textbf{B}\|_F$, respectively. DGCPN-Atanh and DGCPN-HashL only reduce the value gaps between real values and binary codes. As shown in Table \ref{table2}, DGCPN shows the best performance. It is necessary to reduce the similarity gaps between real-valued space and Hamming space, and the half-real and half-binary optimization strategy is an effective approach for obtaining an improved retrieval space.

\begin{figure}[t]
\centerline{
\subfigure[$\alpha$]{\includegraphics[width=0.5\columnwidth]{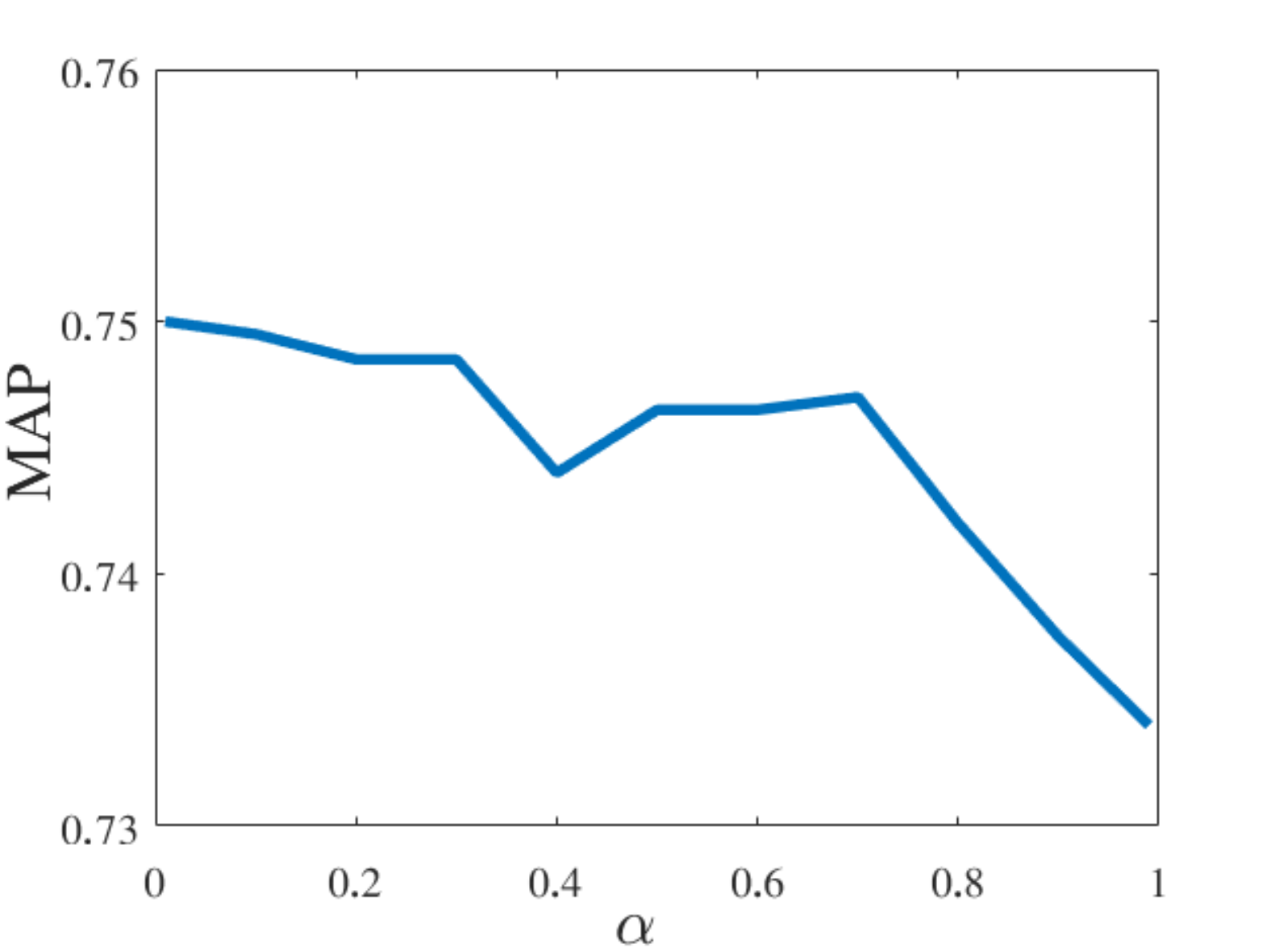}}
\subfigure[$\gamma$]{\includegraphics[width=0.5\columnwidth]{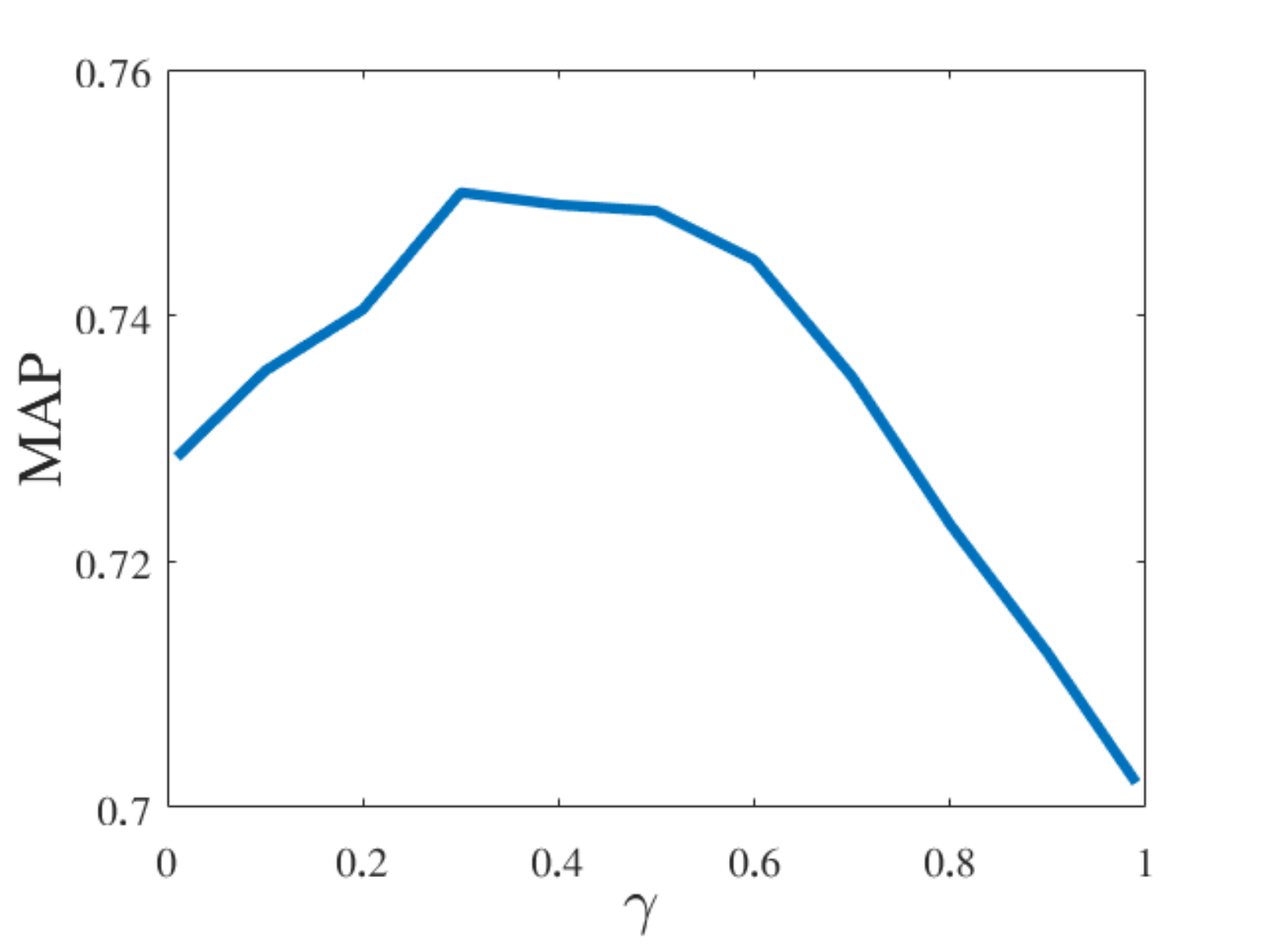}}
}
\centerline{
\subfigure[$k$ and $\beta$]{\includegraphics[width=0.5\columnwidth]{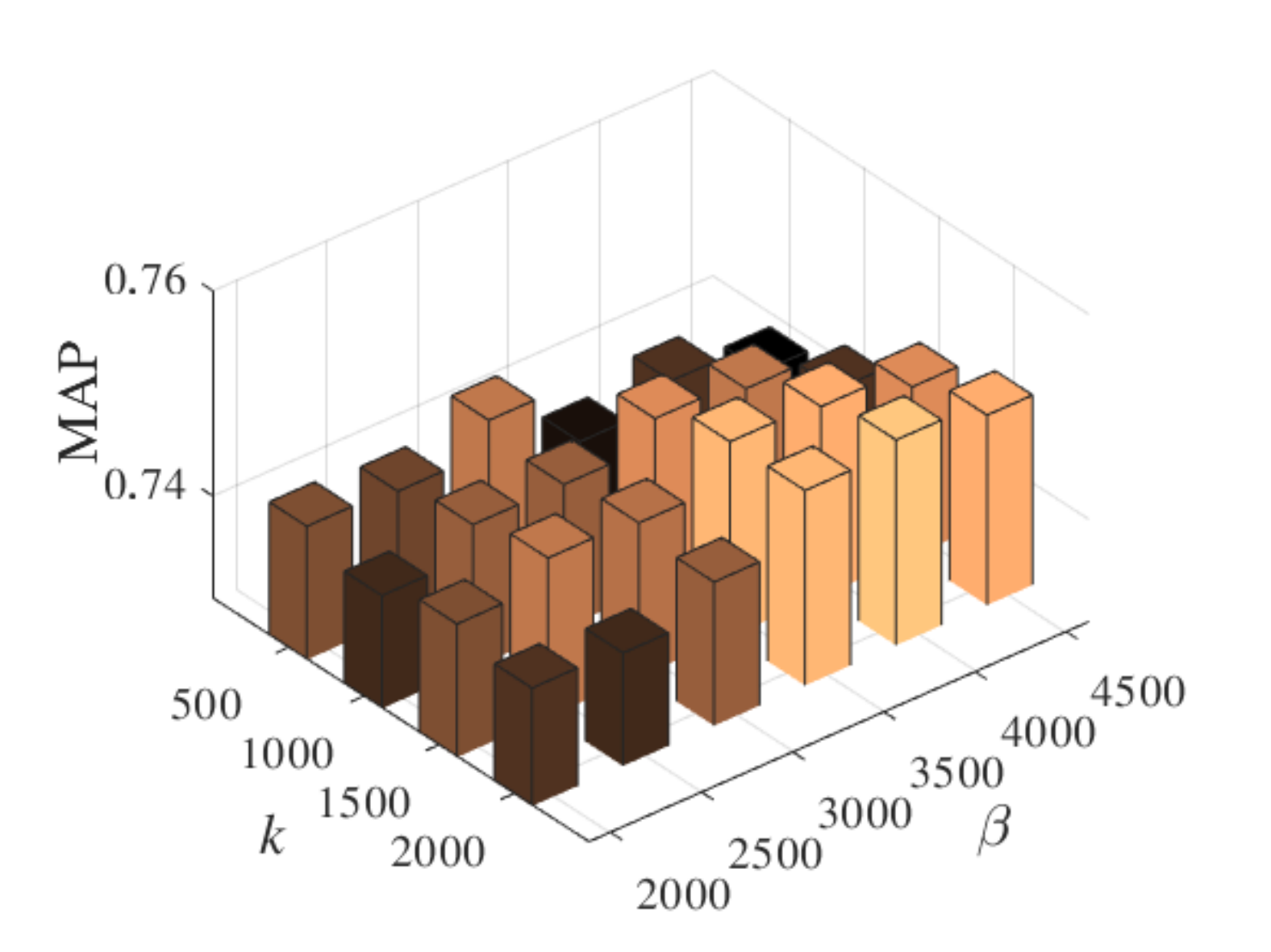}}
\subfigure[$\lambda_1$ and $\lambda _2$]{\includegraphics[width=0.5\columnwidth]{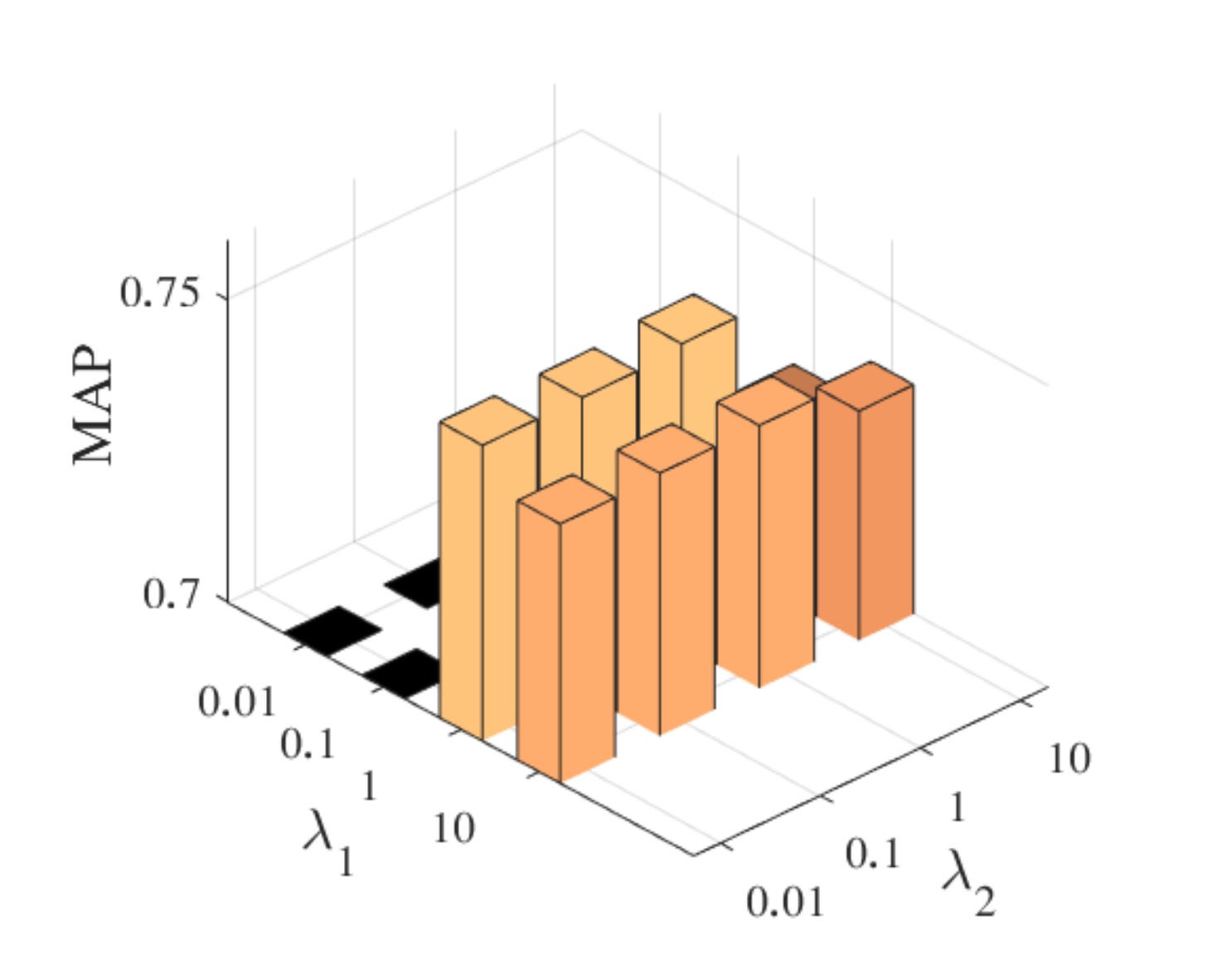}}
}
\caption{Average MAP of the DGCPN
with different parameter values on the MIRFlickr-25K dataset.}
\label{fig4}
\end{figure}

\subsection{Parameter Sensitivity}
Six types of parameters must be adjusted, namely, $\alpha$, $\gamma$, $k$, $\beta$, $\lambda _1$, and $\lambda _2$. In this subsection, we study the performance impacts of these individual parameter values. Fig. \ref{fig4} shows the average MAP of I2T and T2I with different parameter values on the MIRFlickr-25K datasets with the hash code length of 64 bits. From Fig. \ref{fig4}, we can draw the following conclusions: 1) $\alpha$ indicates the relative significance of image and text features. It is observed that a proper combination of cross-modal distances obtains better performance. 2) $\gamma$ adjusts the relative significance of the pairwise distances and graph-neighbor coherence. In Fig. \ref{fig4} (b), we can observe that both higher or lower values of $\gamma$ result in poor performance. 3) $k$ is the number of the $k$-nearest neighbor for each node. $\beta$ is the scale parameter to adjust the range of graph-neighbor coherence. Both parameters contribute to the final value of graph-neighbor coherence. We can see that a proper combination of $k$ and $\beta$ leads to better performance. 4) $\lambda _1$ and $\lambda _2$ are the parameters to adjust the relative significance of the three types of similarity preserving losses. It can be concluded that all three types of similarity preserving losses benefit to the final retrieval performance.

\section{Conclusion}
In this paper, we develop a novel deep graph-neighbor coherence preserving network (DGCPN) for UCMH. Specifically, DGCPN proposes 1) graph-neighbor coherence to explore relationships between data and their neighbors for more accurate data similarities, 2) comprehensive similarity preserving losses to regulate three types of complementary similarities for robust similarity preserving learning, and 3) a half-real and half-binary optimization strategy to reduce value gaps and similarity gaps between the real-valued space and Hamming space for better retrieval performance. The experimental results on three publicly available UCMH datasets compared with the state-of-the-art demonstrate the capability of DGCPN for UCMH.

\textbf{Acknowledgment.} This work was supported in part by the National Nature Science Foundation of China: Grant No. 61836002, No. 62002090, and No. 62020106007, in part by the National Key R$\&$D Program of China: Grant No. 2018AAA0100603, and in part by the Australian Research Council Projects: FL-170100117, DP-180103424, and IH-180100002.

\bibliography{FormattingInstructionsLaTeX2021}
\bibliographystyle{aaai21}
\end{document}